\documentclass{aa}
\usepackage{graphicx}
\begin{document}
   \title{An optical time delay for the 
   double gravitational lens system FBQ 0951+2635\thanks{Based 
   on observations made with the Nordic Optical Telescope, operated 
   on the island of La Palma jointly by Denmark, Finland, Iceland, 
   Norway, and Sweden, in the Spanish Observatorio del Roque de los 
   Muchachos of the Instituto de Astrofisica de Canarias. Based on 
   observations made with ESO Telescopes at the Paranal Observatories 
   under programme ID 66.A-0062(A). Based on observations made with 
   the NASA/ESA Hubble Space Telescope, obtained from the Data Archive 
   at the Space Telescope Science Institute, which is operated by the 
   Association of Universities for Research in Astronomy, Inc., under 
   NASA contract NAS 5-26555. These observations are associated with 
   program \#7887.}}

   \author{P. Jakobsson
          \inst{1,2} \and
          J. Hjorth
          \inst{1} \and
          I. Burud 
          \inst{3} \and
          G. Letawe
          \inst{4} \and
          C. Lidman
          \inst{5} \and
          F. Courbin
          \inst{6}
          }

   \offprints{P. Jakobsson, \\ \email{pallja@astro.ku.dk}}
   
   \institute{Niels Bohr Institute, Astronomical Observatory, 
     University of Copenhagen,
     Juliane Maries Vej 30, DK-2100 Copenhagen \O, Denmark
\and
     Science Institute, University of Iceland, Dunhaga 3, IS-107 
     Reykjav\'{\i}k, Iceland
\and
     Space Telescope Science Institute, 3700 San Martin Drive,
     Baltimore, MD 21218, USA
\and
     Institut d'Astrophysique et de G\'eophysique, ULg, All\'ee du 6
     Ao\^ut 17, B5C, 4000 Sart Tilman (Li\`ege), Belgium
\and
     European Southern Observatory, Casilla 19, Santiago, Chile
\and
     Laboratoire d'Astrophysique, Ecole Polytechnique F\'ed\'erale de
     Lausanne, Observatoire, 1290 Chavannes-des-bois, Switzerland
}
   \date{Received 8 June 2004 / Accepted ???}

   \abstract{
   We present optical $R$-band light curves of the double 
   gravitationally lensed quasar FBQ 0951+2635 from observations 
   obtained at the Nordic Optical Telescope between March 1999 and 
   June 2001. A time delay of $\Delta \tau = 16 \pm 2$\,days (1$\sigma$) 
   is determined from the light curves. New constraints on the lensing 
   geometry are provided by the position and ellipticity of the lensing 
   galaxy. For a $(\Omega_{\mathrm{m}}, \Omega_{\Lambda})= (0.3, 0.7)$ 
   cosmology, the time delay yields a Hubble parameter of $H_0 = 
   60^{+9}_{-7}$~(random, $1 \sigma$)~$\pm 
   2$~(systematic)\,km\,s$^{-1}$\,Mpc$^{-1}$ for a singular isothermal 
   ellipsoid  model and $H_0 = 63^{+9}_{-7}$~(random, $1 \sigma$)~$\pm 
   1$~(systematic)\,km\,s$^{-1}$\,Mpc$^{-1}$ for a constant mass-to-light 
   ratio model. In both models, the errors are mainly due to the time-delay 
   uncertainties. Non-parametric models yield $H_0 = 
   64^{+9}_{-7}$~(random, 
   $1 \sigma$)~$\pm 14$~(systematic)\,km\,s$^{-1}$\,Mpc$^{-1}$.

   \keywords{gravitational lensing -- quasars: individual: FBQ 0951+2635 -- 
   cosmological parameters}
   }

\titlerunning{The gravitational lens system FBQ 0951+2635}

\authorrunning{P. Jakobsson et al.}

\maketitle
\section{Introduction}
Refsdal (\cite{ref}) was the first to point out the possibility of 
determining $H_0$ through gravitational lensing of a variable source. 
Since the light travel times for the various images are unequal, intrinsic 
variations of the source will be observed at different times in the images. 
In particular, the time delay between images is a measurable parameter 
related to the gravitational potential of the lens and $H_0$. There 
is also a weaker dependence on cosmological parameters and on the source 
($z_{\mathrm{s}}$) and lens ($z_{\mathrm{d}}$) redshifts.
\par
Prompted by the successful optical measurements of time delays in QSO 
0957+561 (Schild \& Thomson \cite{schild}; Kundi\'c et al. \cite{kund}) and 
PG 1115+080 (Schechter et al. \cite{pg}), we conducted photometric monitoring 
campaigns of gravitationally lensed quasars at the Nordic Optical Telescope 
(NOT) in La Palma and at the 1.5-m Danish Telescope in La Silla between 1997 
and 2001, yielding four time delays. The first one was obtained for the 
doubly imaged quasar B1600+434 which is lensed by an edge-on spiral (Burud 
et al. \cite{burud}). Then Hjorth et al. (\cite{jens}) estimated the delay 
for the two (summed) components of RX J0911+0550. Finally, delay measurements
were obtained for the doubly lensed quasars HE 2149$-$2745 (Burud et al. 
\cite{burud2}) and SBS 1520+530 (Burud et al. \cite{burud3}). These 
observations increased the total number of firm time-delay measurements 
in gravitational lenses to ten (e.g. Kochanek \& Schechter \cite{ks}).
\par
In this paper we present the fifth result from our programme: 
the time-delay measurement in the doubly-image quasar FBQ 0951+2635. 
It was discovered by Schechter et al. 
(\cite{fbq}, hereafter S98) as a double quasar with angular 
separation of $1\farcs1$. The images are fairly bright with 
mean $R$-band magnitudes of 16.9 and 18.0 for A and B, respectively. The 
time delay between the two images is expected to be relatively short, of 
the order of one month. The monitoring programme, the light curves and the 
time delay are discussed in Sects.~\ref{data.sec}--\ref{time.sec}. 
Spectroscopic observations with the Very Large Telescope (VLT), 
including a measurement of the source redshift and the detection of 
microlensing, are reported in Sect.~\ref{vlt.sec}. In 
Sect.~\ref{mass.sec} we analyse an HST/NICMOS image to constrain a mass 
model of the galaxy and use it to estimate $H_0$. Finally, the main results 
are summarised in Sect.~\ref{dis.sec}.
\section{Observations and data reduction}
\label{data.sec}
\begin{figure}
   \centering
   \resizebox{\hsize}{!}{\fbox{\includegraphics{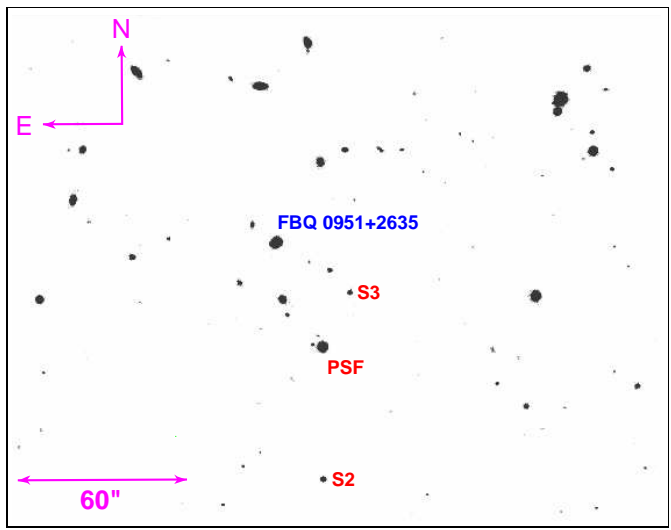}}}
      \caption{A finding chart for FBQ 0951+2635. The star used to model 
      the point spread function (PSF) is indicated. The two reference stars 
      (labeled S2 and S3) used for the photometry are also marked.}
   \label{find.fig}
   \end{figure}

The system was observed on a regular basis at the NOT, starting in 
March 1999. For the first 18\,months it was observed roughly twice per 
month but thereafter around once per week. The target is visible from 
the beginning of October to the end of June, which results in gaps of about 
$\sim$3\,months in the light curves. There were additional gaps due to 
periods of bad weather and bad seeing. Two different instruments were 
used: Andalucia Faint Object Spectrograph and Camera (ALFOSC) and 
the Standby CCD Camera (StanCam), equipped with detectors yielding pixel 
scales of $0\farcs188$ and $0\farcs176$, respectively. In 
Fig.~\ref{find.fig} the central region of the field of view of ALFOSC is 
displayed. One data point 
typically consisted of 1--3 exposures of five minutes each. The seeing 
varied from $0\farcs7$ to $2\farcs8$ during the roughly 
two years of observations, 
with $0\farcs9$ being the most frequent value. All the imaging data were 
pre-processed (bias-subtracted and flat-field corrected) using standard 
IRAF routines.
\section{Photometry}
\label{phot.sec}
The photometric data consist of one stacked frame per epoch. Accurate 
photometry of the blended quasar images is crucial for time-delay 
measurements and all data were analysed with the 
versatile deconvolution algorithm developed by 
Magain, Courbin \& Sohy (\cite{mcs}, hereafter MCS). This algorithm has 
already been used to analyse the data of several blended quasar images 
(e.g. Burud et al. \cite{burud}, \cite{burud2}, \cite{burud3}; Hjorth 
et al. \cite{jens}). The main advantage of using the MCS deconvolution 
in the case of a photometric 
monitoring of a target is its ability to deconvolve all the frames 
from different epochs simultaneously. This constrains the 
positions of the two quasar images, which do not vary with time. 
The intensities, however, are allowed to vary from one image to the other, 
hence producing the light curves.
\par
\begin{figure}
   \centering
   \fbox{\includegraphics[width=4.1cm]{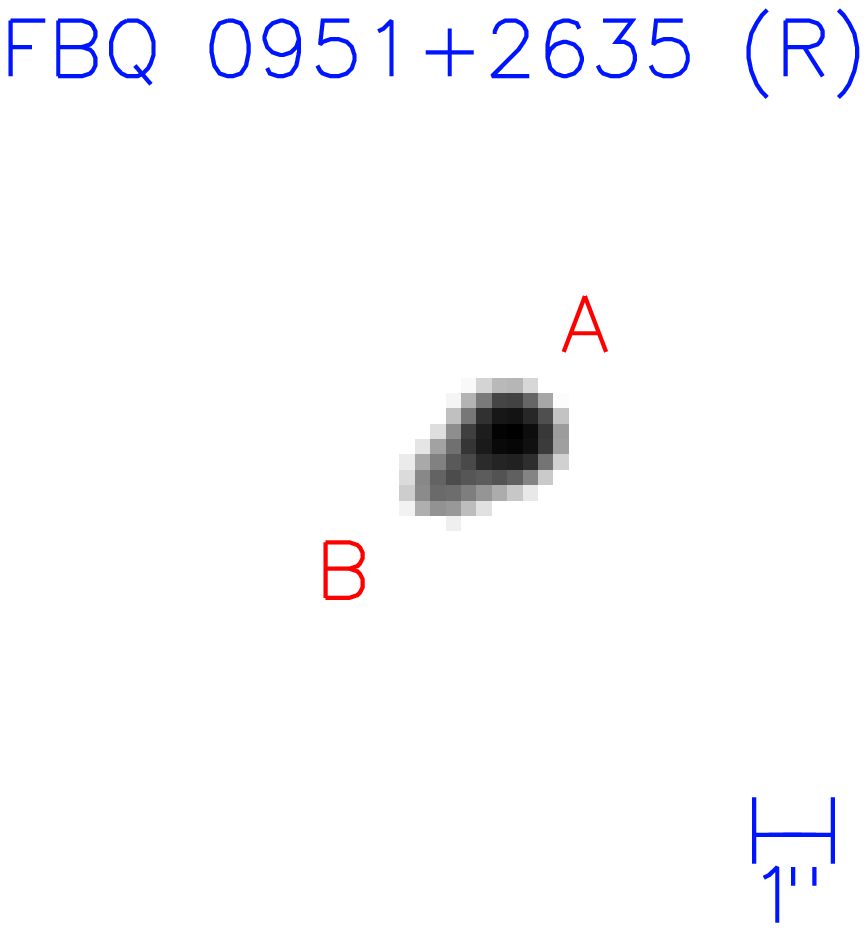}}
   \fbox{\includegraphics[width=4.1cm]{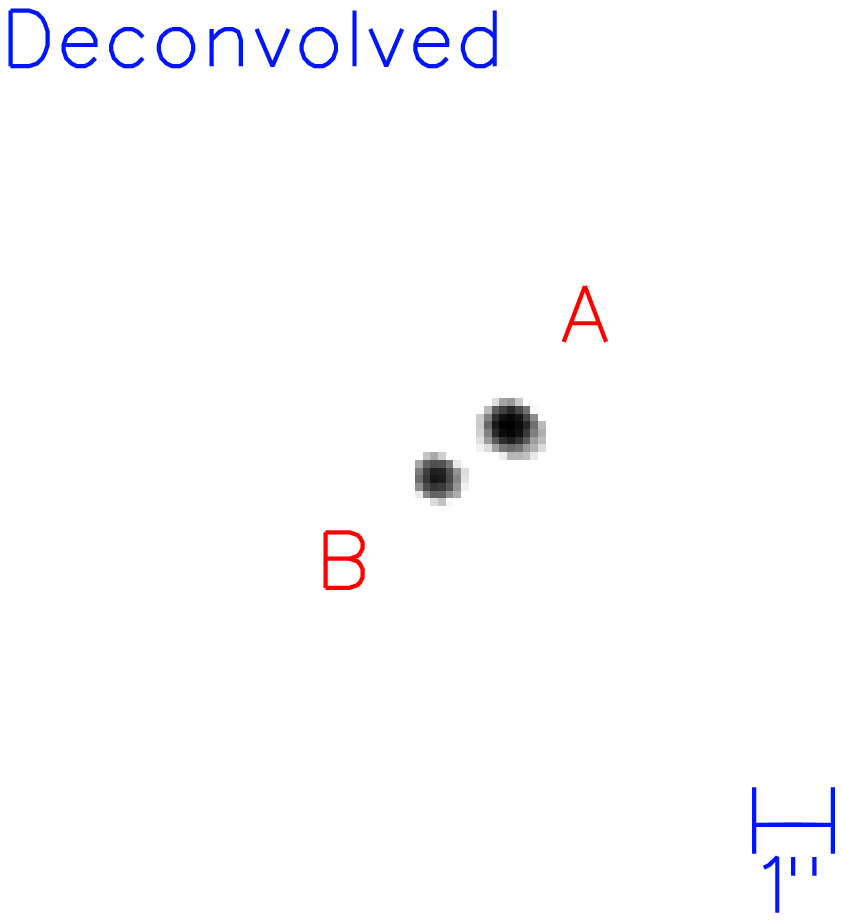}}
      \caption{\emph{Left:} stacked $R$-band image of a 
      $12\arcsec \times 12\arcsec$ region centered on FBQ 0951+2635. 
      The seeing is $\sim$1$\farcs2$ and the total exposure time is 
      $\sim$8.3 hours. \emph{Right:}  deconvolved image 
      (FWHM = $0\farcs28$) obtained from the simultaneous 
      deconvolution of 49 frames. North is up and east to the left.}
   \label{image.fig}
   \end{figure}

The two quasar components are clearly distinguished in our deconvolved image 
of FBQ 0951+2635 (see Fig.~\ref{image.fig}), but the lensing galaxy 
remains too faint to be detected. Although this is clearly a disadvantage 
when measuring the redshift of the galaxy (Sect.~\ref{vlt.sec}), it is 
fortuitous for the time-delay estimate since a contamination by an extended 
object could complicate the analysis.
   \begin{figure}
   \centering
   \resizebox{\hsize}{!}{\includegraphics[bb=75 375 550 695,clip]{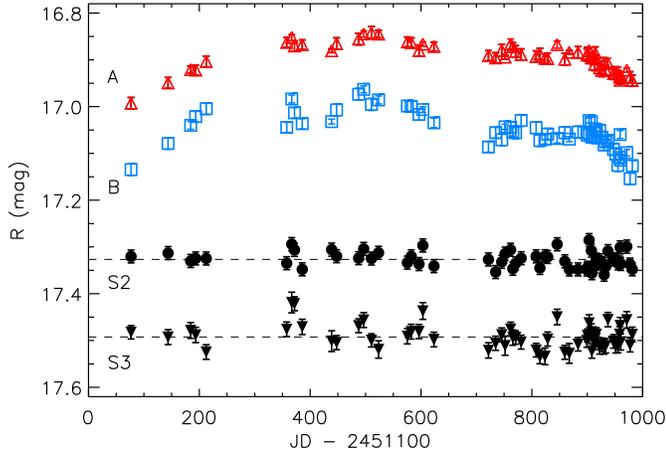}}
      \caption{The A and B component $R$-band light curves for FBQ 0951+2635 
      and two reference stars in the field. The plotted values
      are $R - 0.9$ for B, $R - 1.8$ for S2 and $R - 2.0$ for
      S3. The magnitudes are calculated relative to the calibrated PSF star
      in the field. The error bars represent photon noise and PSF errors
      estimated from the deconvolution of the two reference stars. Since S2 
      and S3 are significantly fainter than components A and B, their photon 
      noise is larger. The dashed horizontal lines represent the mean
      magnitudes of stars S2 and S3.}
   \label{lc.fig}
   \end{figure}
\par
   \begin{figure}[!t]
   \centering
\resizebox{\hsize}{!}{\includegraphics[bb=70 370 560 697,clip]{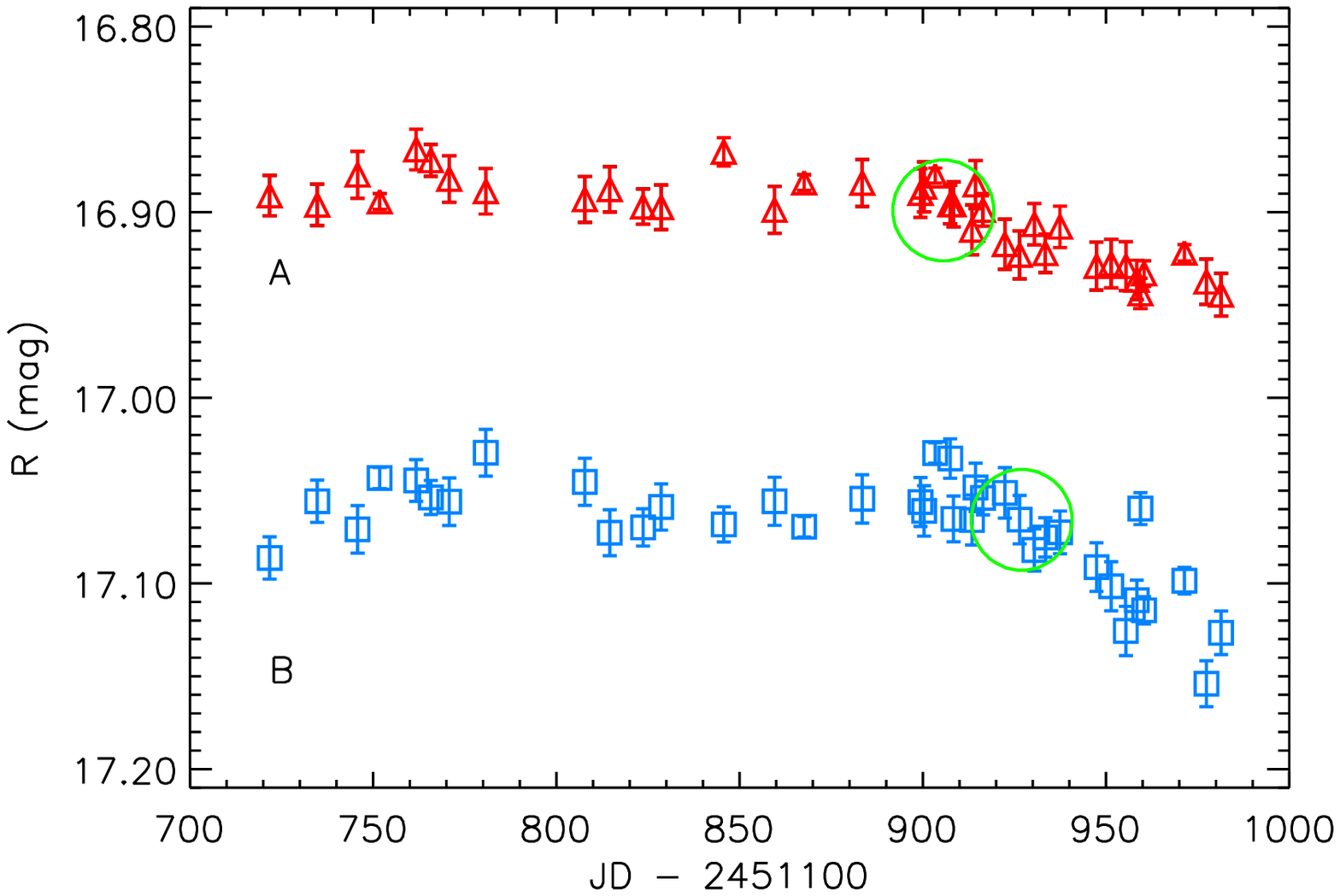}}
\resizebox{\hsize}{!}{\includegraphics[bb=70 370 560 705,clip]{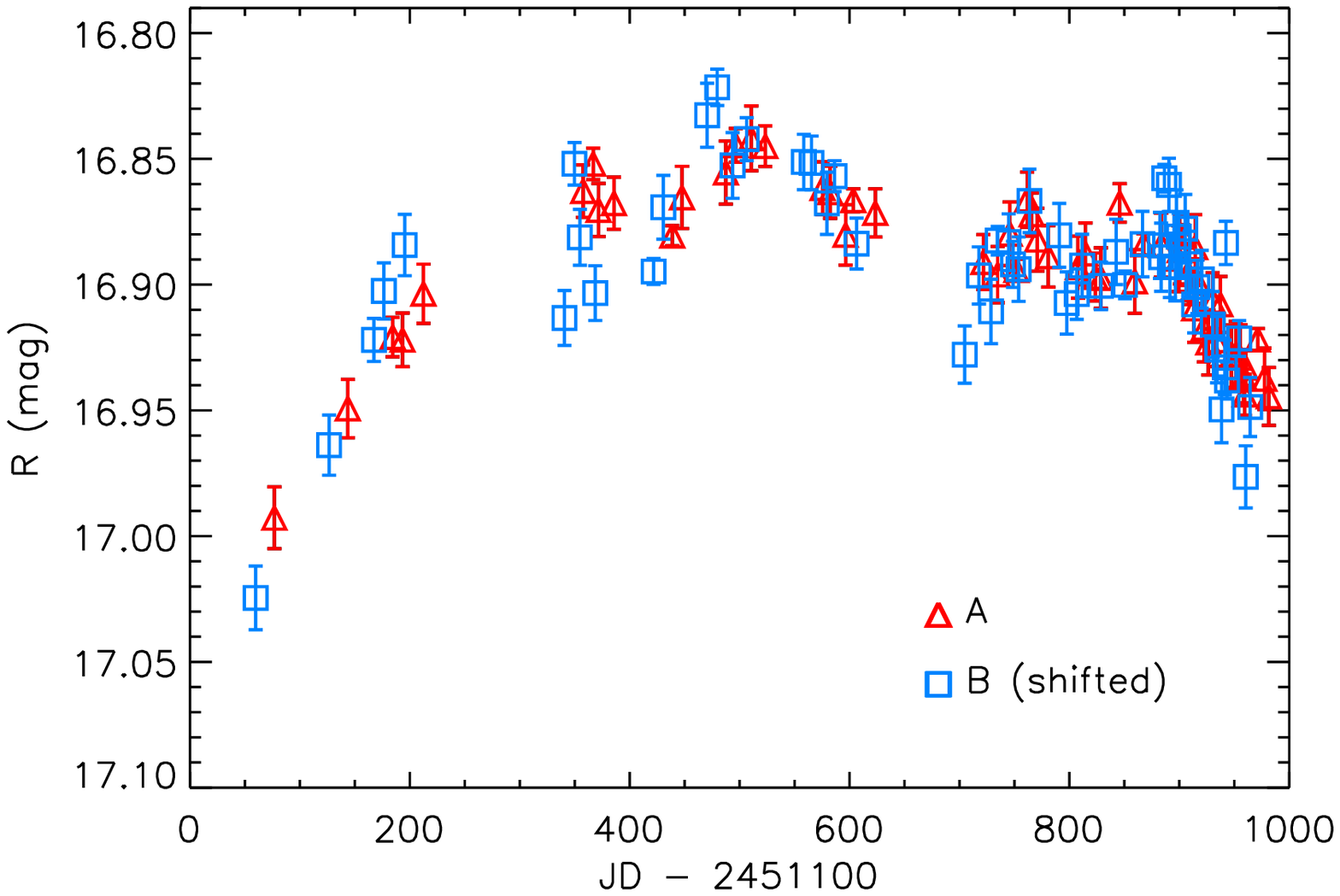}}
\resizebox{\hsize}{!}{\includegraphics[bb=70 370 560 705,clip]{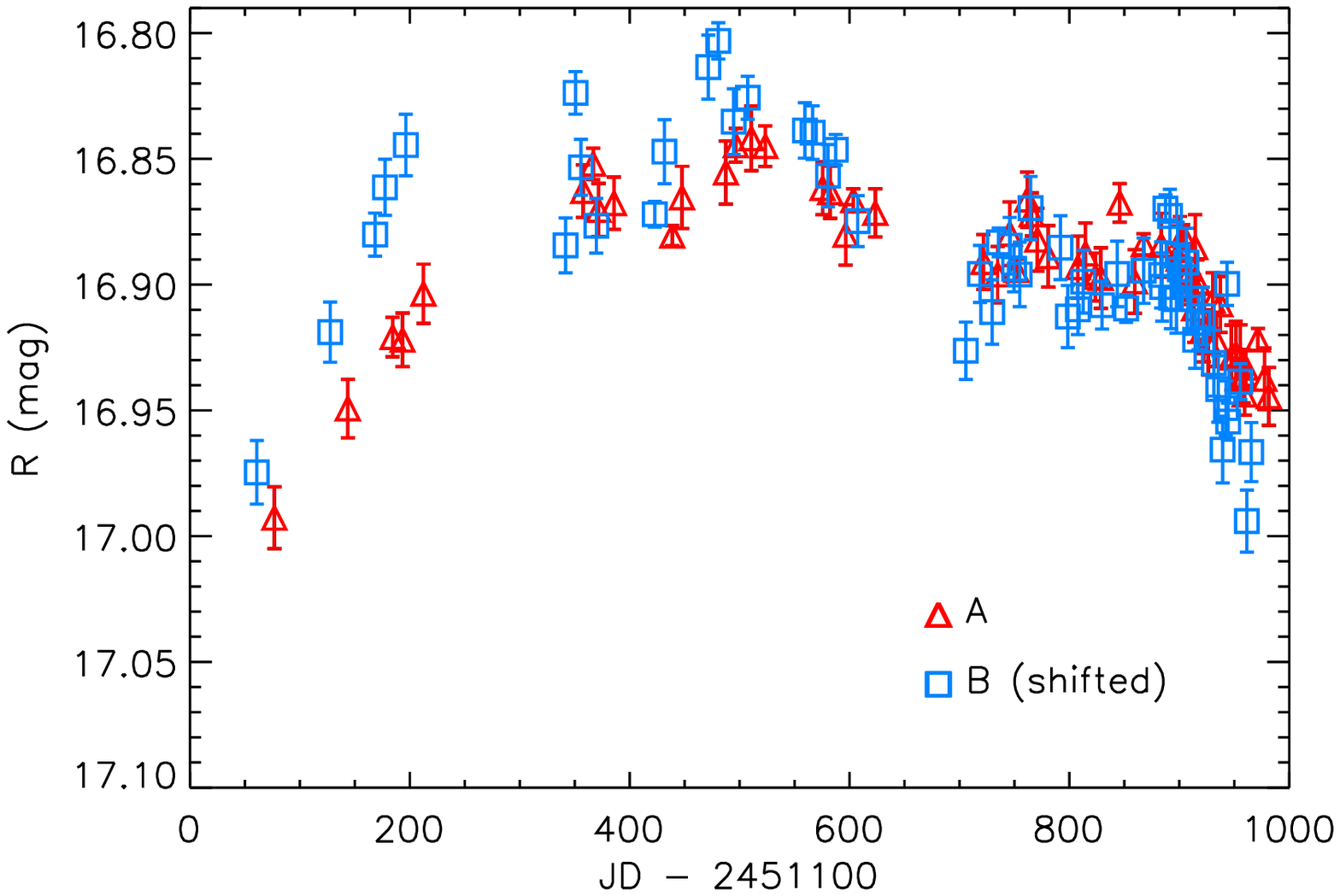}}
      \caption{\emph{Top:} a zoom in on the light curves produced from the 
      latest observations. The plotted values are $R-0.9$ for the B 
      component in order to ease the comparison. A short-lived dip in the 
      light curves is indicated by the two circles. \emph{Middle:} time-delay 
      shifted light curves of FBQ 0951+2635. B is shifted by +1.01 magnitudes 
      and forward in time by 17\,days. A linear correction for the external 
      variations is applied. \emph{Bottom:} time-delay shifted light 
      curves of FBQ 0951+2635. B is shifted by +1.06 magnitudes 
      and forward in time by 16\,days. Only the last 38 data points were 
      used with a direct method (no correction for external variations).}
   \label{lc2.fig}
   \end{figure}
The light curves of FBQ 0951+2635 consist of 58 data points in the 
$R$-band, as presented in Fig.~\ref{lc.fig}. The magnitudes are 
calculated relative to the magnitude of the point spread function (PSF) star 
($R=16.6$) shown in Fig.~\ref{find.fig}. Two other stars were deconvolved 
as well to check for systematic errors. The error bars include both 
photon noise and additional systematic errors, e.g. PSF errors. 
The latter are estimated by using the reference stars, as explained in 
Burud et al. (\cite{burud}).
\section{Time-delay estimate}
\label{time.sec}
It is clear from Fig.~\ref{lc.fig} that the quasar did not show 
large intrinsic variations. But a direct comparison of the first five 
data points of the two components indicates that the time delay is very 
short since the gradual rise in the B component closely follows the rise 
in the A component. In addition, during the last $\sim$100\,days the light 
curves display a nearly simultaneous decrease in flux.
\par
During the observations, a possible tiny short-lived feature was 
detected in both components as shown in the top panel of Fig.~\ref{lc2.fig}. 
Sliding the light curves across one another to find the best visual 
agreement, and making use of this feature, lead us to a rough estimate 
for the time delay of $\sim$15\,days. The light curves show that A is the 
leading component. Once 
light curves of quasar components are obtained, statistical methods are 
needed to determine the time delay more accurately. Using 
the $\chi^2$ minimization method described in Burud et al. 
(\cite{novel}), a more objective value of $\Delta \tau = 17 \pm 4$\,days 
($1 \sigma$) is found from the light curves. The magnitude difference 
between the A and B images is $1.01 \pm 0.01$~mag, corresponding to a
flux ratio between the A and B components of $F_{\mathrm{A}}/F_{\mathrm{B}}
= 2.54 \pm 0.02$. The errors quoted here are obtained from 
Monte Carlo simulations of 1000 sets of light curves, assuming that the 
photometric errors are uncorrelated and follow a Gaussian distribution. 
\par
The method used to reach the aforementioned estimate adds a linear term 
to one of the components, thus modelling slow microlensing effects (in 
Sect.~\ref{vlt.sec} we find independent evidence for microlensing in FBQ 
0951+2635 by analysing its spectra). This linear correction does not 
remove all the external variations as can be seen in the middle panel 
of Fig.~\ref{lc2.fig}. Faster variations on time scales 
of 50--100\,days are still present. The best method for dealing with such 
high-order variations is the iterative version of the algorithm 
(Burud et al. \cite{novel}). It was also applied to the light curves in 
an attempt to correct for these fast variations.
This method yields $\Delta \tau = 13 \pm 4$\,days ($1 \sigma$), 
slightly lower than the value found with the direct method but still in 
agreement within the error bars.
\par
The uncertainties remain large because our method actually finds 
two possible time delays for the whole data set when applying only 
the linear term. In addition to the 17\,days, we also find that 25\,days 
are consistent with the light curves. The result is sensitive to a 
smoothing parameter in the program, which smooths out dates without data 
points on the model curve. Since there are large gaps in our light curves, 
we explore various values for the parameter and find the latter \mbox{result} 
is less robust, with 17\,days a more stable result. To verify this, we also 
ran the programme on the last 38 data points spanning a 260\,day interval 
(shown in the top panel of Fig.~\ref{lc2.fig}), which do not include any 
large gaps and are more frequently sampled . In addition, this interval 
includes the potential short-lived feature. This approach yields a time delay 
of $\Delta \tau = 16 \pm 2$\,days ($1 \sigma$), with no sign of a second 
possible minimum. The magnitude difference between the A and B images is 
$1.06 \pm 0.01$~mag, corresponding to $F_{\mathrm{A}}/F_{\mathrm{B}} = 2.65 
\pm 0.02$. The resulting shifted light curves are displayed in the bottom 
panel of Fig.~\ref{lc2.fig}. In the same figure there are signs of possible 
microlensing variations in one of the components for the first few hundred 
days. This microlensing is most likely the result of compact objects in the 
lensing galaxy (positioned close to the B component; see below), located in
the line-of-sight towards the quasar images. This result is further 
strengthened by the independent spectroscopic microlensing observations in
Sect.~\ref{vlt.sec}.
\begin{table}[t]
\caption[]{Estimated time delays and $R$-band magnitude differences 
for FBQ 0951+2635 calculated with three different methods. The quoted 
uncertainties are $1 \sigma$ errors.}     
\label{time.tab}
\centering
\setlength{\arrayrulewidth}{0.8pt}   
\begin{tabular}{lcc}
\hline
\hline
\vspace{-2 mm} \\
       & $\Delta \tau$ & $\Delta m$ \\
Method & (days)        & (mag)      \\
\vspace{-2 mm} \\
\hline
\vspace{-2 mm} \\
$\chi^2$ fit      &     $17 \pm 4$ & $1.01 \pm 0.01$ \\ 
Iterative fit     &     $13 \pm 4$ & ---             \\ 
Last 38 points    &     $16 \pm 2$ & $1.06 \pm 0.01$ \\ 
\vspace{-2 mm} \\
\hline
\end{tabular}
\end{table}
\par 
The time-delay estimates and magnitude offsets obtained from our three 
different approaches are consistent with each other and are presented in 
Table~\ref{time.tab}. Given that the large gaps in our light curves affect 
our results, we adopt $\Delta \tau = 16 \pm 2$~days as the best estimate 
for the time delay. 
\section{VLT spectroscopy}
\label{vlt.sec}
The redshift of the lensing galaxy has proved elusive. S98 found absorption 
lines attributed to \ion{Mg}{ii} at $z=0.73$ and $z=0.89$ in both spectra 
of the quasar components. From the position of the lens on the fundamental 
plane, Kochanek et al. (\cite{konni}) suggested $z_{\mathrm{d}}=0.21$ with a 
range of 0.18--0.23, while Kochanek et al. (\cite{konni2}, hereafter K98) 
quote a redshift of 0.24, estimated from the properties of the lensing 
galaxy (C. Kochanek 2002, private communication). Multiply lensed quasars 
have also become an important tool to detect microlensing by compact 
objects within the lensing galaxy. This offers the opportunity to study 
the spatial structure of the lensed quasar with very high angular resolution.
\par
On 2001 March 24--25 we obtained deep optical spectra of FBQ 0951+2635 
with the VLT using the Multi Object Spectroscopy (MOS) capability of 
FORS1 with $1\farcs4$ wide slitlets. Two grisms were used, 
G300V+GG435 covering approximately 4300--8600\,\AA\ and G300I+GG435 with 
a wavelength coverage of roughly 6600--11000\,\AA. 
\par
We acquired $2 \times 1200 + 620$\,s spectra with G300V and 
$3 \times 1200$\,s spectra with G300I. The slitlets were 
oriented with a position angle of $-55\fdg6$ to cover both
quasar components and the lensing galaxy. One of the slitlets was
centred on a star in the field of view whose PSF was later used as an 
input for deconvolution of the quasar spectra. The 
individual spectra were combined, yielding a seeing of $0\farcs4$ 
($0\farcs6$) and a spectral resolution of 10\,\AA\ (16\,\AA) FWHM 
for G300I (G300V). The projected pixel size was $0\farcs1 \times 
2.59$\,\AA\ ($0\farcs1 \times 2.69$\,\AA) for G300I (G300V).
\par
Redshift estimations of lensing galaxies are often limited by observational
difficulties arising from the fact that the light from these galaxies can
be heavily blended with the light from one of the quasar images, the
latter sometimes being several magnitudes brighter. The spectral 
deconvolution algorithm developed by Courbin et al. (\cite{courbin})
is able to deconvolve spectra in the spatial direction. It can 
therefore be used to separate the quasar and galaxy spectra. 
This technique was applied to FBQ 0951+2635 in order to deblend the spectra 
of the two quasar images and to search for the lensing galaxy, located only 
$0\farcs2$ away from component B.
\par
In the spectra of both quasar components we identify three emission lines 
at $>$$5\sigma$, namely \ion{C}{iii}] $\lambda 1909$, \ion{Mg}{ii} 
$\lambda 2798$ and [\ion{O}{ii}] $\lambda 3727$. These lines place the 
quasar at $z_{\mathrm{s}} = 1.246 \pm 0.001$, a value fully consistent
with that found by White et al. (\cite{white}). In addition, the absorption 
lines first reported by S98, interpreted as the \ion{Mg}{ii} doublet 
at $z=0.729$ and $z=0.892$, are detected in the A component. In the spectrum 
of the B component we only detect the $z=0.892$ system.
\par
Strong equivalent width (EW) difference exists between the \ion{Mg}{ii} line 
in the spectra of the two quasar images, with EW a factor of 1.46 smaller in 
component B than in component A with high significance. The most probable 
explanation is that the continuum emission of the B component is 
micro\-lensed by compact objects in 
the lensing galaxy. This result is consistent with standard models of quasar 
structure, with the scale of the broad emission line region significantly 
greater than that of the continuum source (e.g. Wambsganss et al. \cite{wamb}; 
Krolik \cite{krolik}). The \ion{Mg}{ii} flux ratio between A and B is 
$3.7 \pm 0.3$. This is considerably higher than the continuum flux ratio 
($\sim$2.5) and that from broadband photometry.
\par
In order to obtain the lensing galaxy spectrum, free of any contamination 
by the two bright nearby quasar images, the 2-D spectrum of FBQ 0951+2635 was 
spatially deconvolved. However, the signal-to-noise ratio was very low in the 
lensing galaxy spectrum. Due to this and a lack of significant absorption 
and emission lines in the spectrum, we were unable to determine a 
spectroscopic redshift. In what follows we adopt $z_{\mathrm{d}} = 0.24$ 
from K98, but note that decreasing $z_{\mathrm{d}}$ from 0.26 to 0.18 
decreases the value of $H_0$ by $37\%$.
   \begin{figure*}[t]
   \centering
   \fbox{\includegraphics[width=8.65cm]{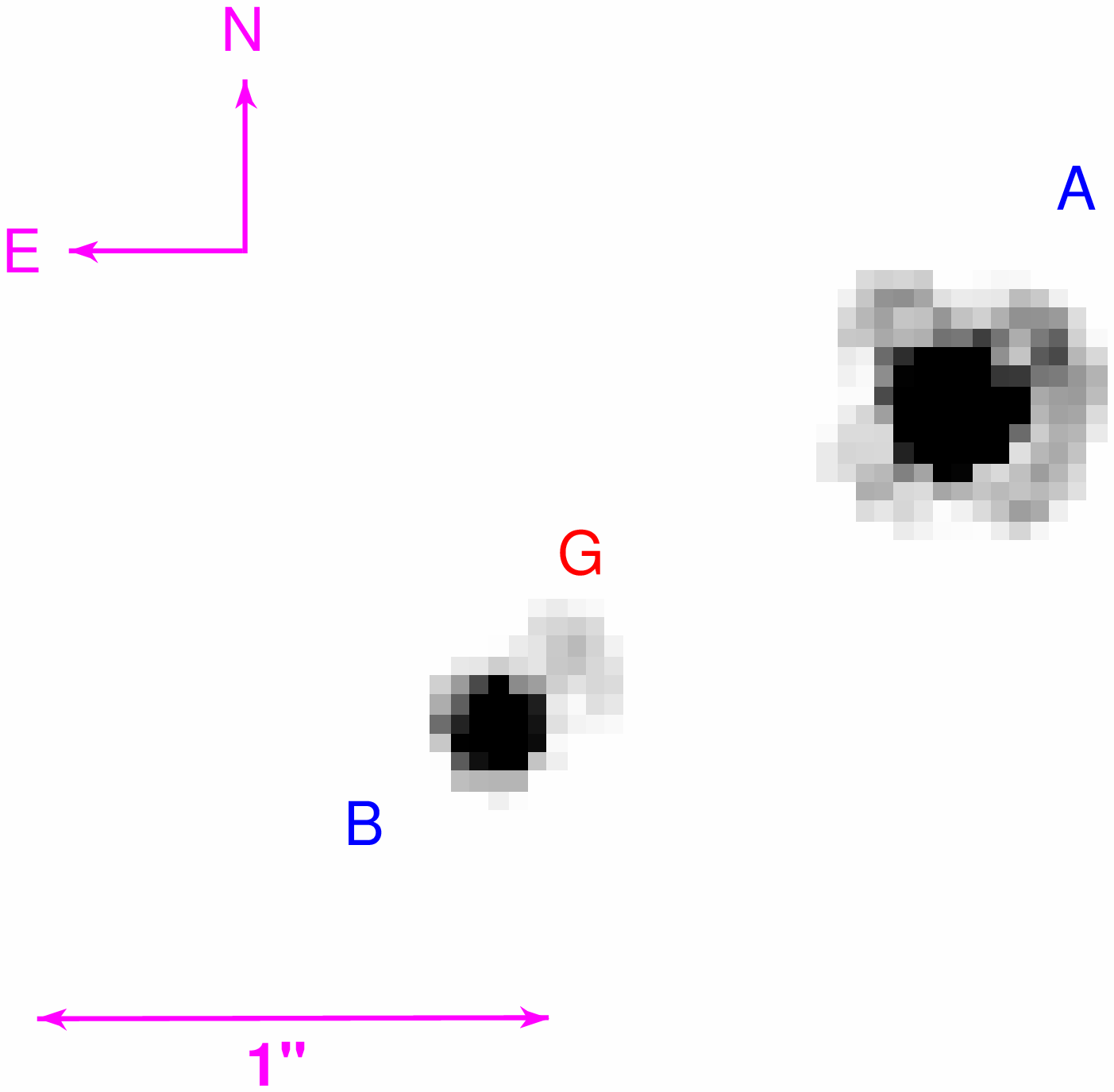}}
   \fbox{\includegraphics[width=8.65cm]{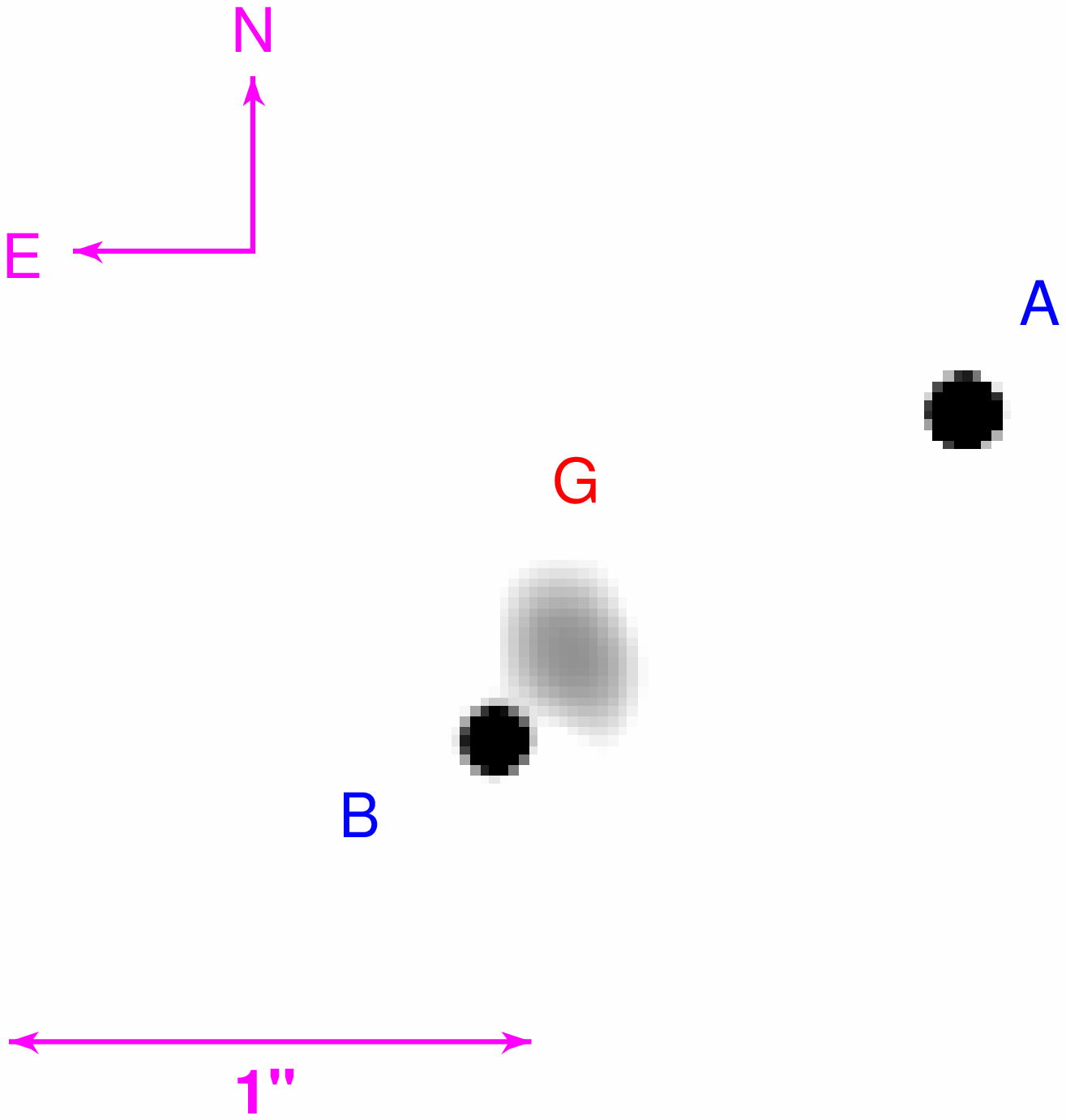}}
      \caption{\emph{Left:} an HST/NICMOS image of FBQ 0951+2635. 
      \emph{Right:} a deconvolved HST/NICMOS image shown on a logarithmic
      intensity scale. In both panels the lensing galaxy is denoted by the 
      letter 'G'.}
   \label{ella.fig}
   \end{figure*}
\section{Lens modelling}
\label{mass.sec}
The time delay and redshifts measured for FBQ 0951+2635 can be 
used to infer an estimate of $H_0$, based on modelling of the total 
gravitational potential responsible for the lensing effect. This includes 
the main lensing galaxy and any tidal perturbations from objects near it or 
along the line of sight. FBQ 0951+2635 is not an easy case to model, as 
the lensing galaxy is not readily visible. However,
as seen in the left panel of Fig.~\ref{ella.fig}, there is a small 
'lightbump' northwest of, but close to the B component. 
\par
We use two distinct approaches to convert the time delay into $H_0$: 
parametric models that involve an analytical form for the lensing 
galaxy, and non-parametric models that involve more degrees of freedom, 
i.e. a pixelated surface density (Saha \& Williams \cite{sw97}; Williams \& 
Saha \cite{ws}; Saha \& Williams \cite{sw04}).
\subsection{HST/NICMOS results}
\label{nicmos.sec}
In March 1998, four 640\,s images of the FBQ 0951+2635 lens system were 
obtained by K98, taken with the Near Infrared 
Camera and Multi-Object Spectrometer's (NICMOS) NIC2 camera onboard the
HST. The filter used was F160W. The images were drizzled to give a 
pixel scale of $0\farcs0375$.
\par
We deconvolved the drizzled image, using the A component as an empirical 
PSF. The right panel of Fig.~\ref{ella.fig} shows the result from the 
deconvolution: the lensing galaxy is positioned close to the B component. 
This lens configuration is in full agreement with the fact that A is the 
leading component (see Sect.~\ref{time.sec}), as predicted by gravitational
lens theory. We fit ellipses to the isophotes in the core of the lensing 
galaxy resulting in an ellipticity of $e = 0.25 \pm 0.04$ and a position angle 
(PA) of $22^{\circ} \pm 4^{\circ}$. Assuming that the mass distribution 
follows the light distribution we can restrict ourselves to these ranges of 
ellipticities and PAs. For the de Vaucouleurs model (see below) we also 
estimated the effective radius, resulting in $R_{\mathrm{e}} = 0\farcs09 
\pm 0\farcs02$. The exact positions of the quasar images are listed in 
Table~\ref{pos.tab} in a lens-centered coordinate system. The quasar images 
have an angular separation of $1\farcs091 \pm 0\farcs001$, while the B 
component is positioned $0\farcs221 \pm 0\farcs002$ from the center of the 
lensing galaxy. We note that our position of the galaxy obtained from 
the HST/NICMOS image is quite different from that published by S98, where 
a lens candidate is detected midway between the two components. But the S98 
data were ground-based and they acknowledge that the candidate might be a
spurious detection.
\begin{table}
\caption[]{Image and lens positions.}     
\label{pos.tab}
\centering
\setlength{\arrayrulewidth}{0.8pt}   
\begin{tabular}{lrr}
\hline
\hline
\vspace{-2 mm} \\
Object & $\theta_x$ \hspace{0.7cm} & $\theta_y$ \hspace*{0.7cm}\\
\vspace{-2 mm} \\
\hline
\vspace{-2 mm} \\
Lens & $0\farcs000 \pm 0\farcs002$ & $0\farcs000 \pm 0\farcs002$ \\
A & $0\farcs750 \pm 0\farcs001$ & $0\farcs459 \pm 0\farcs001$ \\
B & $-0\farcs142 \pm 0\farcs001$ & $-0\farcs169 \pm 0\farcs001$ \\
\vspace{-2 mm} \\
\hline
\end{tabular}
\end{table}
\subsection{Analytical models}
Using the \emph{lensmodel} package (Keeton \cite{chuck}), we fit
the lens system with a dark matter model and a constant $M/L$ model. 
The former is a singular isothermal ellipsoid (SIE) model with a flat 
rotation curve, while the latter is the prototypical constant $M/L$ model: 
the de Vaucouleurs model. This is a similar approach as Kochanek \& Schechter
(\cite{ks}) carried out for four lens systems with known time delays 
measured by Burud et al. (\cite{burud}, \cite{burud2}, \cite{burud3}) and 
Schechter et al. (\cite{pg}). The SIE model sets a lower bound on $H_0$, 
while the constant $M/L$ model sets an upper bound on $H_0$. This results 
from the fact that no galaxy in a cold dark matter dominated Universe should 
have mass distributions which are more centrally concentrated than the 
luminosity distribution (Mo et al. \cite{mmw}). In addition, a typical
lensing galaxy should not have a rising rotation curve at approximately 
twice the effective radius, making the SIE the least concentrated reasonable
model.
\par
During the fit, the astrometry of the quasar images relative to the lens 
is fixed, as well as the redshift of the source and lens. The models are 
constrained by using the range of ellipticities and PAs obtained
in Sect.~\ref{nicmos.sec}. In addition, we use $F_{\mathrm{A}}/F_{\mathrm{B}}$ 
as a constraint. Four values have been derived independently. In 
Sect.~\ref{time.sec} we found from broadband photometry that 
\mbox{$F_{\mathrm{A}}/F_{\mathrm{B}} = 2.5$--2.7}. This is similar to the 
ratio calculated from the spectral continua (Sect.~\ref{vlt.sec}). On the 
other hand, the emission line flux ratio is measured to be 
$F_{\mathrm{A}}/F_{\mathrm{B}} = 3.7 \pm 0.3$ (Sect.~\ref{vlt.sec}). Finally, 
S98 report the radio emission line ratio to be 
$4.7 \pm 0.6$. While flux ratios from broadband photometry are generally known 
to be unreliable due to the ever present possibility of microlensing, the
emission line fluxes should reflect the ``true'' flux ratio (e.g.
Wisotzki et al. \cite{wis}). Hence, in what follows, we
adopt $F_{\mathrm{A}}/F_{\mathrm{B}} = 3.7 \pm 0.3$ as a model constraint,
but note that allowing $F_{\mathrm{A}}/F_{\mathrm{B}}$ to vary between 
2.5 and 4.7 does change the value of $H_0$ by only approximately $4\%$.
 \begin{figure}
 \centering
 \resizebox{\hsize}{!}{\includegraphics[bb=75 370 536 699,clip]{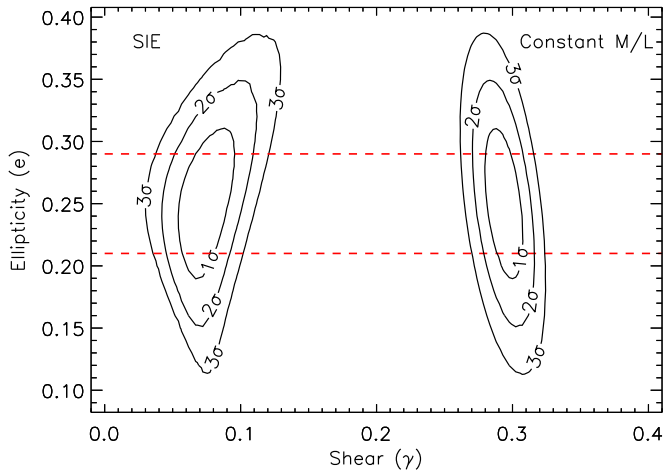}}
    \caption{$\Delta \chi^2$ contours for FBQ 0951+2635 in the plane of
    the ellipticity ($e$) and shear ($\gamma$). The contours are drawn at 
    $\Delta \chi^2$ = 2.30, 6.17 and 11.8, corresponding to the $1 \sigma$, 
    $2 \sigma$ and $3 \sigma$ confidence intervals for two degrees of freedom. 
    There is a weak degeneracy between $e$ and $\gamma$. The area between 
    the horizontal dashed lines is the range of the measured ellipticity.}
  \label{chi2.fig}
  \end{figure}
\par
The free parameters in our model include the mass of the lensing galaxy, 
the source coordinates and flux, and the amplitude of the external shear 
($\gamma$) along with the direction to it. We include shear in our models 
since at least two mass concentrations lie along the line of sight, namely 
the ones responsible for the \ion{Mg}{ii} absorption systems (S98; 
Sect.~\ref{vlt.sec}).
It is important to understand any degeneracies between the two sources of 
angular structure in the model, the ellipticity and shear. The external 
shear breaks the circular symmetry of the lens and therefore it has the 
same effect as introducing ellipticity in the lens. Figure~\ref{chi2.fig} 
shows the contours of the $\Delta \chi^2$ in the space of $e$ and 
$\gamma$ after optimizing all other parameters. In the SIE model this 
results in a moderate shear of $\gamma = 0.07$, while for the constant $M/L$ 
model there is a large shear of $\gamma = 0.29$. By evaluating $H_0$ inside 
the $1 \sigma$ contours in Fig.~\ref{chi2.fig} we get an estimate for the 
systematic uncertainties in both models. If we take the currently most 
popular cosmology, i.e. $\Omega_{\mathrm{m}} = 0.3$, $\Omega_{\Lambda}=0.7$, 
and $\Delta \tau = 16$~days, we find $H_0 = 60^{+9}_{-7}$~(random, 
$1 \sigma$)~$\pm 2$~(systematic)\,km\,s$^{-1}$\,Mpc$^{-1}$ for the
SIE model, while $H_0 = 63^{+9}_{-7}$~(random, $1 \sigma$)~$\pm 
1$~(systematic)\,km\,s$^{-1}$\,Mpc$^{-1}$ for the constant $M/L$ model.
Due to the relatively low redshift of the lensing galaxy, the influence 
of a change in cosmology is small compared to the measurement errors. 
Adopting other cosmologies, 
$(\Omega_{\mathrm{m}}, \Omega_{\Lambda})=(1.0, 0.0)$ or 
$(\Omega_{\mathrm{m}}, \Omega_{\Lambda})=(0.3, 0.0)$, will change the inferred 
value of $H_0$ by only about $4\%$ for this lens system.
\subsection{Pixelated models}
While the mass distribution of the pixelated models of Saha \& Williams 
(\cite{sw04}) is not as well related to the physical parameters of the lens,
the models allow an exploration of a wide range of lens shapes. Thus,
this approach better constrains the systematic errors as it is unrestricted
by the confines of the parametric models.
\par
 \begin{figure}
 \centering
 \resizebox{\hsize}{!}{\includegraphics[bb=68 360 542 705,clip]{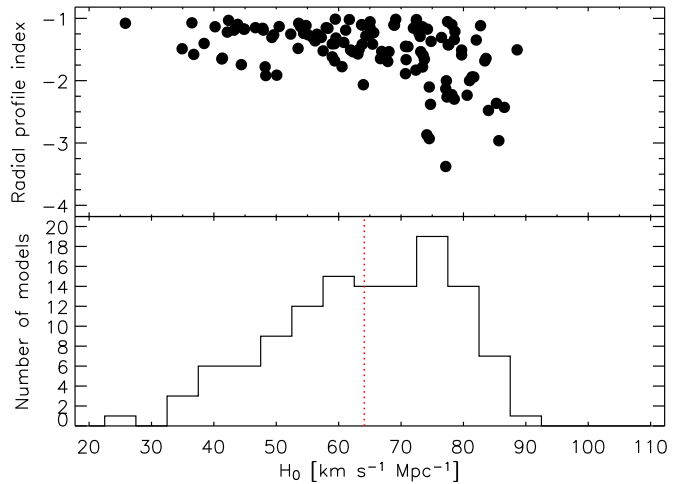}}
    \caption{\emph{Top:} the slope of the projected mass density profile 
     plotted against $H_0$ for 121 different non-parametric models.
     \emph{Bottom:} probability distribution of the derived $H_0$ 
     values for the same models as above. The median value of the distribution 
     is indicated by the dotted line.}
  \label{pixelens.fig}
  \end{figure}
We fit the lens system with the \emph{PixeLens} 
programme.\footnote{\scriptsize{http://ankh-morpork.maths.qmw.ac.uk/$\sim$saha/astron/lens/pix/}}
During the fit, the ellipticity of the mass distribution was kept free. 
As in the analytical models, we added external shear. The
lensing galaxy appears unperturbed in Fig.~\ref{ella.fig}, leading us to
enforce the inversion symmetry constraint (i.e. the lens looks the same
if rotated by $180^{\circ}$). In order to avoid models less concentrated 
than the SIE, we imposed a constraint stating that the index of the projected
radial mass density profile ($\alpha$) must be steeper than $-1$. As there is 
not an option in \emph{PixeLens} to invoke this constraint, we ran a total of 
500 models and, in the upper panel of Fig.~\ref{pixelens.fig}, plot those
121 that fulfill $\alpha < -1$. Most lens systems, when modelled 
independently, show a correlation between $\alpha$ and $H_0$, in the sense
that steeper density profiles result in higher values of $H_0$. FBQ 
0951+2635 shows this correlation rather weakly. The lower panel of 
Fig.~\ref{pixelens.fig} displays the derived $H_0$ probability distribution. 
The mean of the distribution is calculated to be 
$H_0 = 64^{+9}_{-7}$~(random, 
$1 \sigma$)~$\pm 14$~(systematic)\,km\,s$^{-1}$\,Mpc$^{-1}$.
\section{Discussion}
\label{dis.sec}
The time delay $\Delta \tau = 16 \pm 2$\,days ($1 \sigma$) has been measured 
in the lensed quasar FBQ 0951+2635 on the basis of $R$-band images obtained 
with the NOT. Due to the lack of pronounced features in the light curves and 
because the time delay is short compared to the sampling of the light curves,
$\Delta \tau$ is not very well constrained, thus explaining the large error 
bars. If this short time delay is to be pinned down more accurately, 
intensive monitoring should be performed in the future. As long as the 
variability is sufficiently strong and erratic during the observations, 
this should help constrain $\Delta \tau$ even further.
\par
Applying an SIE lens model to the case of FBQ 0951+2635, we derive 
$H_0 = 60^{+9}_{-7}$~(random, $1 \sigma$)~$\pm 
2$~(systematic)\,km\,s$^{-1}$\,Mpc$^{-1}$, where the error
bars incorporate the uncertainty in the measured time delay as well as
the uncertainties in the lens model due to the degeneracy between the
shear and ellipticity. For a constant $M/L$ model we estimate 
$H_0 = 63^{+9}_{-7}$~(random, $1 \sigma$)~$\pm 
1$~(systematic)\,km\,s$^{-1}$\,Mpc$^{-1}$. These two analytical models 
were studied because they represent the limiting case for physically
possible mass distributions given our current understanding of dark
matter. Non-parametric models employed on the lens system resulted in
$H_0 = 64^{+9}_{-7}$~(random, 
$1 \sigma$)~$\pm 14$~(systematic)\,km\,s$^{-1}$\,Mpc$^{-1}$, with 
a relatively broad range in $\alpha$.
\par
Unfortunately we cannot determine which mass profile is appropriate
given the available data. However, by using an independent $H_0$
estimate we can try to break the degeneracy. The local estimate by 
the HST Key Project (Freedman et al. \cite{freedman}) is 
$H_0 = 72 \pm 8$\,km\,s$^{-1}$\,Mpc$^{-1}$, the recent results by WMAP
(Spergel et al. \cite{wmap}) indicate that $H_0 =
71^{+4}_{-3}$\,km\,s$^{-1}$\,Mpc$^{-1}$, while the value of 
$H_0 = 59 \pm 6$\,km\,s$^{-1}$\,Mpc$^{-1}$ by Saha et al. (\cite{saha}) is 
based on Cepheid calibration for a sample of galaxies with Type Ia
supernovae. Due to our large $\Delta \tau$ uncertainty, all three
estimates are consistent with the models we adopt. It is therefore
important to monitor this lens system more frequently in order to beat
down the random uncertainty on any $H_0$ estimate from FBQ 0951+2635.
\begin{acknowledgements}
     It is a pleasure to thank Chuck Keeton for helpful discussions about 
     lens modelling. We thank the anonymous referee for his/her critical
     reading and useful comments on the paper. We are especially grateful 
     to Anlaug Amanda Kaas and the many visiting observers at NOT who have 
     contributed to this project by performing the scheduled observations. 
     We thank Gunnlaugur Bj\"ornsson and Einar Gudmundsson for constructive 
     comments and suggestions. PJ gratefully acknowledges support from a 
     special grant from the Icelandic Research Council. FC acknowledges 
     financial support from the European Commission through Marie Curie grant 
     MCFI-2001-0242 and the P\^ole d'Attraction Interuniversitaire, P5/36 
     (PPS Science Policy, Belgium). Some of the optical 
     data presented here have been taken using ALFOSC, which is owned by the 
     Instituto de Astrofisica de Andalucia (IAA) and operated at the Nordic 
     Optical Telescope under agreement between IAA and the NBIfAFG of the 
     Astronomical Observatory of Copenhagen. This work was supported by 
     the Danish Natural Science Research Council (SNF).
\end{acknowledgements}

\end{document}